\documentclass[twocolumn,showpacs,preprintnumbers,amsmath,amssymb]{revtex4}

\usepackage{graphicx}
\usepackage{dcolumn}
\usepackage{bm}

\begin{document}

\title{Magnetic coupling properties of rare-earth metals (Gd, Nd)
doped ZnO: first-principles calculations}

\author{Hongliang Shi$^1$}
\email{hlshi@semi.ac.cn}
\author{Ping Zhang$^2$}
\author{Shu-Shen Li$^1$}
\author{Jian-Bai Xia$^1$}
 \affiliation{$^1$State Key Laboratory for
Superlattices and Microstructures, Institute of Semiconductors,
Chinese Academy of Sciences, Beijing 100083, People's Republic of
China \\$^2$LCP, Institute of Applied Physics and Computational
Mathematics, P.O. Box 8009, Beijing 100088, China}

\begin{abstract}
The electronic structure and magnetic coupling properties of
rare-earth metals (Gd, Nd) doped ZnO have been investigated using
first-principles methods. We show that the magnetic coupling between
Gd or Nd ions in the nearest neighbor sites is ferromagnetic. The
stability of the ferromagnetic coupling between Gd ions can be
enhanced by appropriate electron doping into ZnO:Gd system and the
room-temperature ferromagnetism can be achieved. However, for ZnO:Nd
system, the ferromagnetism between Nd ions can be enhanced by
appropriate holes doping into the sample. The room-temperature
ferromagnetism can also be achieved in the \emph{n}-conducting
ZnO:Nd sample. Our calculated results are in good agreement with the
conclusions of the recent experiments. The effect of native defects
(V$_{\rm{Zn}}$, V$_{\rm{O}}$) on the ferromagnetism is also
discussed.
\end{abstract}

\maketitle

\section{\label{sec:level1}INTRODUCTION}
The diluted magnetics semiconductors (DMSs) such as 3\emph{d}
transition metals (TMs) doped ZnO have attracted a lot of attention
due to their great potential applications in the spintronic devices
\cite{r100,r200,r300,r400,r1,r2,r201,r3,r301,r4,r401}. Recently,
rare earth (RE) ions doped DMSs have also invoked great interests
since the colossal magnetic moment of Gd in GaN was reported by Dhar
\emph{et al.} \cite{r5}. They reported that the average value of the
moment per Gd is as high as 4000 $\mu_{B}$ which can be explained in
terms of a long-range spin polarization of the GaN matrix by Gd
\cite{r5}. First principles studies have been carried out by
different teams to explain ferromagnetism in GaN:Gd \cite{r6,r7}.
Dalpian and Wei \cite{r6} reported that the coupling between Gd
atoms is antiferromagnetic and the electrons can stabilize the
ferromagnetic phase because of the coupling between the Gd \emph{f}
and host \emph{s} states introduced by the same symmetry. However,
Liu \emph{et al.} \cite{r7} argued that the room-temperature
ferromagnetism in GaN:Gd can be explained by the interaction of Gd
4\emph{f} spins via \emph{p-d} coupling involving holes introduced
by intrinsic defects and holes are more effective than electrons in
contributing to the observed colossal magnetic moment of Gd ions.

Experimental studies for the ferromagnetism of ZnO:Gd systems have
also produced controversial conclusions \cite{r8,r9}. Potzger
\emph{et al.} \cite{r8} found ferromagnetism in Gd-implanted ZnO
single crystals. They noted that when sufficient density of Gd ions
is present, then annealing is necessary to release enough charge
carriers to establish the ferromagnetic coupling of the diluted Gd
ions. However, Ungureanu \emph{et al.} \cite{r9} reported that there
are no exchange interaction between the RE ions and a large negative
magnetoresistance is obtained which can be interpreted as a
paramagnetic response of the system to the applied magnetic field.

Compared with 3\emph{d} transition metals (TMs), 4\emph{f} rare
earth (RE) metals have larger magnetic moments. Furthermore, the
electrons may mediate the ferromagnetic coupling between the RE
ions, due to the coupling between \emph{f} electrons (with
\emph{a$_{1}$} symmetry) and host \emph{s} electrons \cite{r6}. This
conclusion may be good for ZnO based DMSs, because grown ZnO films
or single crystals always exhibit \emph{n}-type conductivity. Until
now, no theoretical studies about RE ions doped ZnO systems are
found. Therefore, it is interesting to investigate the electronic
structures and magnetic couplings properties of ZnO:RE.

In this work, we have systematically studied the magnetic coupling
between the RE ions with different electron and hole concentrations
doped into the ZnO:RE samples. We find that the coupling between Gd
ions in ZnO is ferromagnetic. Furthermore, the electrons of
appropriate concentration can enhance the ferromagnetic coupling
between them. However, for Nd ions, the holes of appropriate
concentration can enhance the ferromagnetic coupling between them.
The effect of native defects (V$_{\rm{Zn}}$, V$_{\rm{O}}$) on the
ferromagnetism is also discussed. We note that although the oxygen
vacancy can contribute electrons to the system, the coupling between
Gd ions is anti-ferromagnetic. Maybe this is because the
concentration of electrons introduced by oxygen vacancy is too high
in our studied supercell.

\section{\label{sec:level1}DETAILS OF CALCULATION}
Our first-principles calculations are based on the density
functional theory (DFT) and the Vienna ab initio simulation package
\cite{r10} using the generalized gradient approximation (GGA) of PBE
functional \cite{r11} for the exchange correlation potential. The
electron and core interactions are included using the frozen-core
projected augmented wave (PAW) approach \cite{r12}. The Zn 3\emph{d}
and rare earth metals 5\emph{p}4\emph{f} electrons are explicitly
treated as valence electrons. The electron wave function is expanded
in plane waves up to a cutoff energy of 400 eV. For the Brillouin
zone integration, a 2$\times$2$\times$4 Monkhorst-Pack
\emph{k}-point mesh is used for the supercell containing 32 atoms
and a good convergence is obtained. All the geometries are optimized
until the quantum mechanical forces acting on the atoms are smaller
than 0.01 eV/$\rm{\AA}$. In our present work, we calculate the
ferromagnetic properties for the RE ions doped in the zinc-blende
structure ZnO and we expect our conclusions are similar to those for
RE ions doped in ground phase ZnO with wurtzite structure since the
band structures of the ZB and WZ alloy are very similar near the
band edge at $\Gamma$ \cite{r14}. In our magnetic calculations, we
substitute two Zn atoms with two RE ions in the nearest neighbor
(NN) sites, corresponding to a concentration of 12.5\% for RE ions.

\section{\label{sec:level1}RESULTS AND DISCUSSIONS}
\subsection{\label{sec:level1}Electronic structure of hypothetical zinc-blende phase GdO and NdO}
In order to obtain a clear understanding of the magnetic coupling of
the RE ions, we first study the electronic structures of the
hypothetical zinc-blende (ZB) phase GdO and NdO binary alloys. The
present calculated lattice constants for ZB structure ZnO, GdO, and
NdO are 4.626, 5.375, and 5.444 \AA, respectively. Our calculated
results show that both GdO and NdO are more stable in the
ferromagnetic phases and the energy differences between
anti-ferromagnetic (AFM) and ferromagnetic (FM) coupling states are
16 meV and 291 meV, respectively.

Based on the crystal field theory, in a tetragonal substitutional
site of ZB structure, the \emph{d} orbitals are split into one
triply degenerate \emph{t$_{2}$} state and one doubly generate
\emph{e} state; the \emph{f} orbitals are split into two triply
degenerate \emph{t$_{2}$}, \emph{t$_{1}$} states and one singly
\emph{a$_{1}$} state. The spin-resolved band structures of GdO and
NdO are plotted in Fig. 1 and Fig. 2, respectively. The labels in
these two figures represent the characters of Bloch wave functions
at $\Gamma$ point.

In Fig. 1, due to the large electronegativity of oxygen, the
4\emph{f} majority spin channels are above the oxygen \emph{p}
states, unlike the GdN in which the 4\emph{f} majority spin channels
are below the nitrogen \emph{p} states \cite{r6}. At $\Gamma$ point,
the O \emph{p} states with \emph{t$_{2}$} symmetry in spin up
channels are below those in spin down channels. This is because at
$\Gamma$ point the coupling between the Gd \emph{t}$_{2f}$ and O
\emph{t}$_{2p}$ states is larger in the spin up channel, which
pushes down the O \emph{p} states. We also notice that at $\Gamma$
point, the Gd \emph{s} state is located below the Fermi energy in
both spin up and spin down channels. Furthermore, for GdN, Gd is
isovalent with Ga because of the
4\emph{f$^{7}$}5\emph{d$^{1}$}6\emph{s$^{2}$} valence configuration,
therefore, GdN is semiconducting \cite{r6}. For GdO, however, Gd is
not isovalent with Zn, thus GdO is not semiconducting. From the
density of states (DOS) of GdO plotted in Fig. 3(a), we can also
conclude that GdO is not semiconducting because there are majority
spin states at the Fermi-energy position and the density of states
are not symmetrical for the majority and minority spins.

In Fig. 2, the band structure of the ferromagnetic phase NdO with ZB
structure is showed.  At $\Gamma$ point, part of the 4\emph{f}
states are occupied in the spin up channel below the Fermi energy,
while the 4\emph{f} states in the spin down channel are fully empty.
According to the DOS plotted in Fig. 3(b), we see that the 4\emph{f}
bands of majority spin are located between -0.5 eV and 0.4 eV, and
the peak in the DOS of 4\emph{f} states of minority spin occurs
around 2.3 eV. Overall, the DOS of 4\emph{f} states are broadened
due to the \emph{f-s} hybridization.

\begin{figure}
\includegraphics*[height=8cm,keepaspectratio]{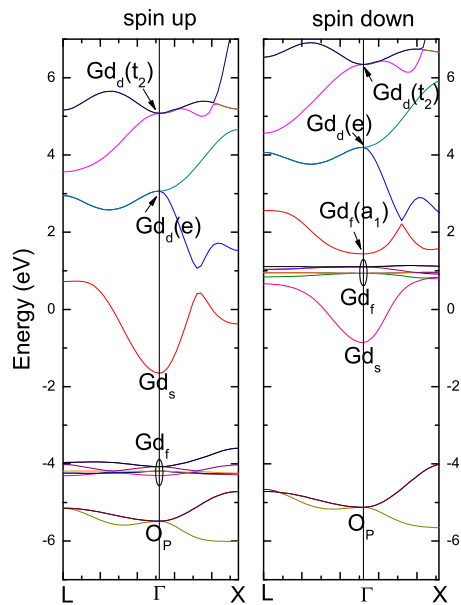}
\caption{\label{fig:epsart}Band structure of the ferromagnetic phase
GdO with ZB structure. The labels represent the symmetry character
of the band at $\Gamma$ point.}
\end{figure}

\begin{figure}
\includegraphics*[height=8cm,keepaspectratio]{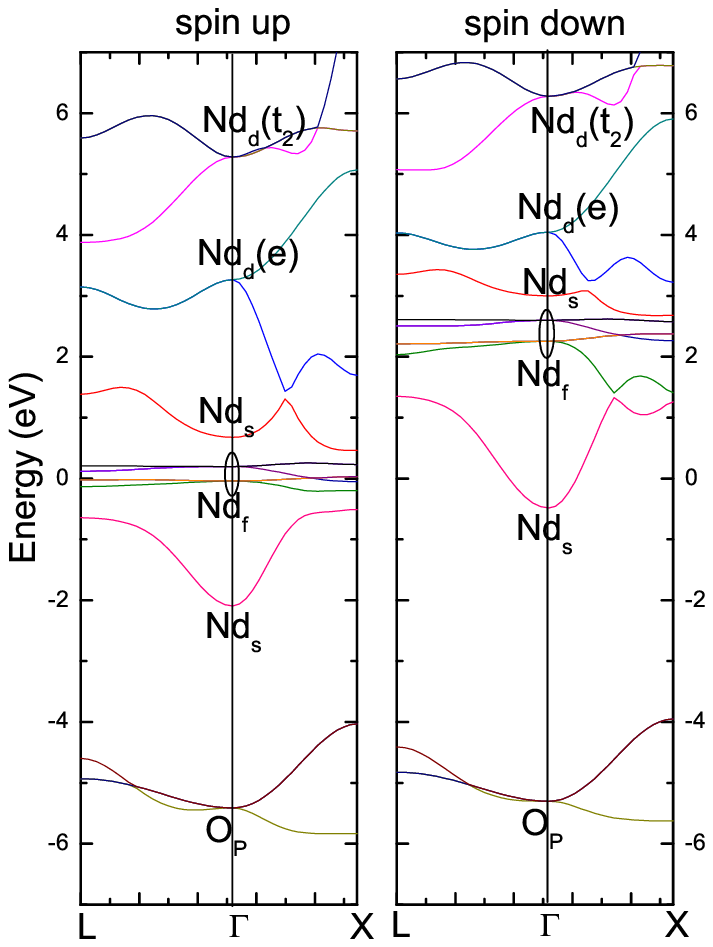}
\caption{\label{fig:epsart}Band structure of the ferromagnetic phase
NdO with ZB structure. The labels represent the symmetry character
of the band at $\Gamma$ point.}
\end{figure}

\begin{figure}
\includegraphics*[height=6.6cm,keepaspectratio]{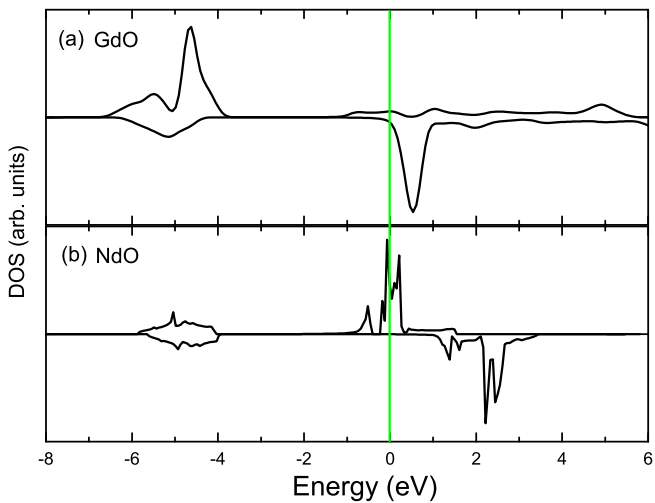}
\caption{\label{fig:epsart}Density of states of (a) GdO and (b) NdO.
The Fermi energy is taken to be zero and the green line is to make
it more clear.}
\end{figure}

\subsection{\label{sec:level1}Ferromagnetic coupling of Gd and Nd ions doped in ZnO}

\begin{figure}
\includegraphics*[height=8cm,keepaspectratio]{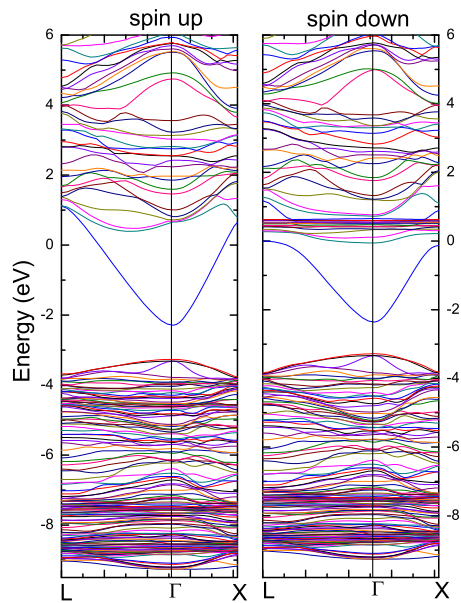}
\caption{\label{fig:epsart}Band structure of ferromagnetic
Zn$_{14}$Gd$_{2}$O$_{16}$}
\end{figure}

\begin{figure}
\includegraphics*[height=8cm,keepaspectratio]{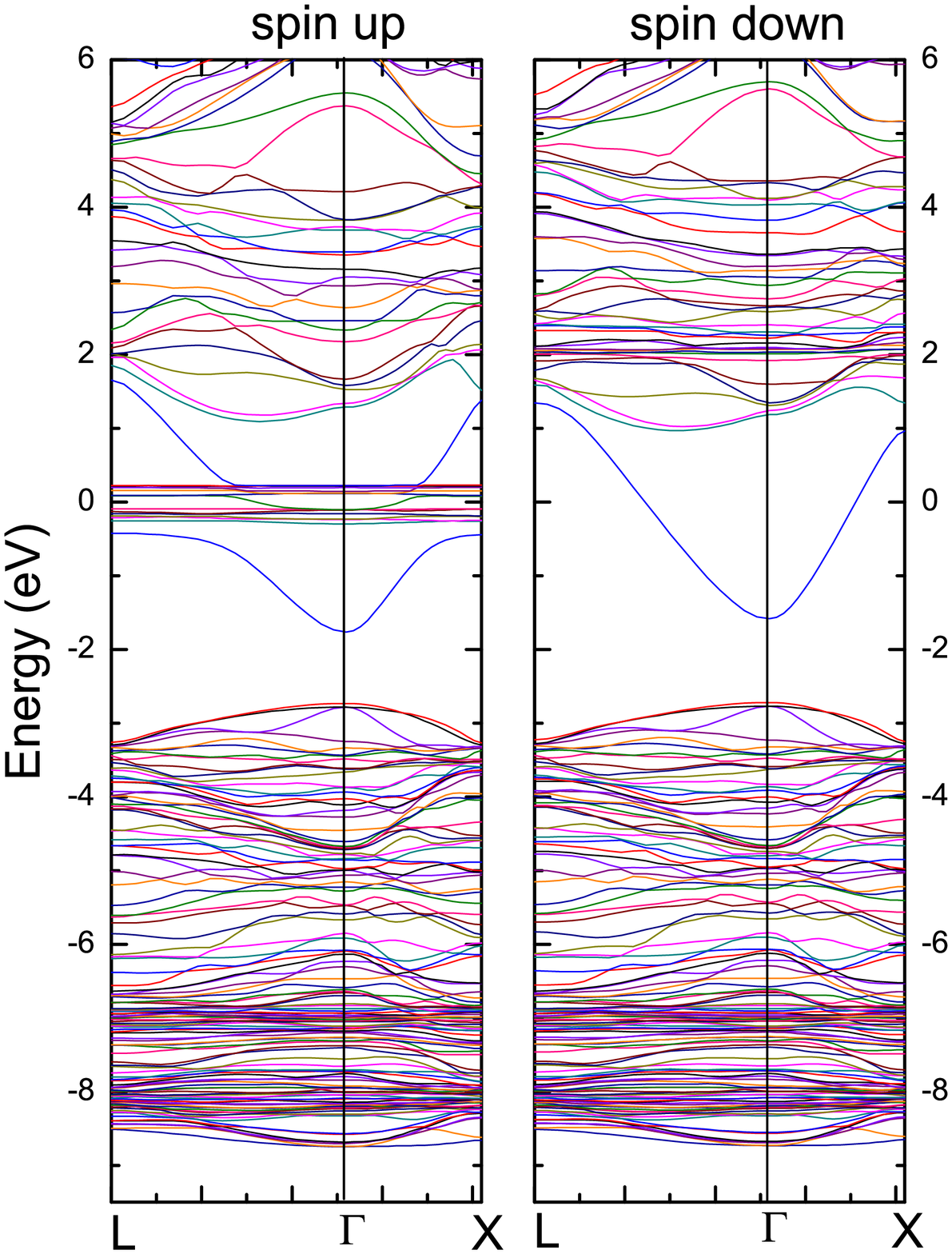}
\caption{\label{fig:epsart}Band structure of ferromagnetic
Zn$_{14}$Nd$_{2}$O$_{16}$}
\end{figure}

\begin{table}
\caption{\label{tab:table1}Calculated energy differences $\Delta$E
($\Delta$E=E$_{\rm{AFM}}$-E$_{\rm{FM}}$) between anti-ferromagnetic
and ferromagnetic configurations for ZnO:Gd and ZnO:Nd in different
charged states.}
\begin{ruledtabular}
\begin{tabular}{lcc}
RE & charge state & $\Delta$E (meV)\\
 \hline
    & neutral &  7\\
    & 0.15e/Gd & 44\\
    & 0.25e/Gd & 67\\
    & 0.50e/Gd & 61\\
Gd  & 0.75e/Gd & -8\\
    & 1.00e/Gd   &-37\\
    & 0.50h/Gd & 4 \\
    & 1.00h/Gd   & 0 \\
    & V$_{\rm{Zn}}$ & 0 \\
    & V$_{\rm{O}}$ & -8\\
    \hline
    & neutral &  94\\
    & 0.25h/Nd & 128   \\
    & 0.50h/Nd & 134\\
    & 0.75h/Nd & 69\\
Nd  & 1.00h/Nd & 20  \\
    & 0.25e/Nd & 46 \\
    & 0.50e/Nd & 16 \\
    & V$_{\rm{Zn}}$ & 52 \\
    & V$_{\rm{O}}$ & 67\\
\end{tabular}
\end{ruledtabular}
\end{table}

In the following, we systematically study the properties of
ferromagnetic coupling between Gd (Nd) ions in different charged
states. Experimentally, Potzger \emph{et al.} \cite{r8} found
ferromagnetism in Gd-implanted ZnO single crystals. They further
reported that when sufficient density of Gd ions is present, then
annealing is necessary to release enough charge carriers to
establish the ferromagnetic coupling of the diluted Gd ions. In our
present study, we substitute two Zn atoms with two RE ions (Gd or
Nd) in the nearest neighbor sites. We simulate the effect of donors
(acceptors) by introducing electrons (holes) into the ZnO:RE
systems. We want to know how the electrons and holes with different
concentrations mediate the ferromagnetism in the ZnO:RE systems. We
have also investigated whether the intrinsic defects (V$_{\rm{Zn}}$,
V$_{\rm{O}}$) play an important role in the magnetic properties of
ZnO:RE systems. Our results show that the ground states of Gd and Nd
doped ZnO systems are all ferromagnetic coupling states in the
neutral state, and the energy differences $\Delta$E between the
anti-ferromagnetic coupling state (AFM) and the ferromagnetic
coupling state (FM) are 7 and 94 meV, respectively. The direct
coupling between 4\emph{f} electrons is very weak compared with the
energy difference $\Delta$E for 3\emph{d} transition metals doped in
ZnO \cite{r4,r401,r18}. This is because the orbitals of 4\emph{f}
electrons are very localized \cite{r6}. The band structure of FM
phase for ZnO:Gd and ZnO:Nd systems are plotted in Fig. 4 and Fig.
5, respectively. In Fig. 4 for ZnO:Gd system, the 4\emph{f} bands of
the majority spin (lying at about $-20$ eV) are not plotted because
they are far from the Fermi energy. However, in Fig. 5 for ZnO:Nd
system, the 4\emph{f} bands of the majority spin are located near
the Fermi energy due to the partially occupied 4\emph{f} orbitals.

Due to the \emph{s-f} coupling, electrons may mediate the
ferromagnetism in \emph{n}-type samples \cite{r16,r17,r18}.
Experimentally, Ungureanu \emph{et al.} \cite{r19} also reported
that the presence of ferromagnetism in ZnO:Gd films might indicate
electron mediated interion exchange. In order to test this
speculation, we insert different amounts of electrons into the
ZnO:Gd supercell. Remarkably, we find that even when 0.15 electron
per Gd is inserted into the ZnO:Gd system, the stability of the FM
phase is enhanced with $\Delta$E=44 meV
($\Delta$E=E$_{\rm{AFM}}$-E$_{\rm{FM}}$), much larger than the
neutral case of $\Delta$E=7 meV. We also calculate the cases of
0.25, 0.50, 0.75, and 1.0 electron per Gd. The corresponding
$\Delta$E are 67, 61, $-8$, and $-37$ meV. We note that the ZnO:Gd
system will favor AFM phase if 0.75 or 1.0 electron per Gd is
inserted into the system. In practice, however, even for 0.5
electron per Gd, such high doping corresponding to 1.314 $\times$
10$^{21}$ cm$^{-3}$ is hard to achieve experimentally. For hole
doping, the calculated $\Delta$E is 4 meV after creating 0.5 hole
per Gd in the system, thus indicating that the stability of the FM
phase is weakened by the hole doping. In particular, when the hole
doping is as high as 1.0 hole per Gd, we find that the energies of
FM and AFM phases of ZnO:Gd are degenerate and thus the system
favors paramagnetic phase. Since the intrinsic defects are important
as well for the magnetic properties, we have also investigated how
the zinc and oxygen vacancies affect Gd-Gd magnetic interaction. For
zinc vacancies V$_{\rm{Zn}}$, in our studied supercell one
V$_{\rm{Zn}}$ contributes 2 holes to the system, and our calculated
$\Delta$E is 0 meV, i.e., the ZnO:Gd system favors paramagnetic
phase. This is consistent with the case of 1.0 hole/Gd, both can
lead the ZnO:Gd system to favor paramagnetic phase. However, one
V$_{\rm{O}}$ introduces two electrons to the system and leads the
system to anti-ferromagnetic phase with $\Delta$E=$-$8 meV. The
effect of one V$_{\rm{O}}$ on the system is similar to that of the
case of 1.0 electron per Gd.

Our above calculated results are listed in table I. According these
results we can conclude that electrons can effectively mediate the
ferromagnetism of the ZnO:Gd system through controlling its
concentrations. Ungureanu \emph{et al.} \cite{r19} also reported
that the presence of ferromagnetism at 300 K in ZnO:Gd films with
around 1\% Gd co-doped with 0.2\% Al might indicate electron
mediated exchange in ZnO:Gd systems. In their samples, Al servers as
\emph{n}-type dopant. The concentration of electrons introduced by
Al (1\% Gd co-doped with 0.2\% Al) is in the range of our studied
cases of 0.15$\sim$0.25 electron per Gd, which, as predicted from
Table I, also leads the ZnO:Gd system to favor ferromagnetic phase.
As for the Curie temperature (\emph{T$_{c}$}), our calculated energy
differences $\Delta$E in Table I suggest that for the present
calculated ZnO:Gd system with the electron concentration comparable
with that in attainable experiment \cite{r19}, the derived
\emph{T$_{c}$} is higher than room temperature (RT). This
prediction, which is really consistent with the experiment
\cite{r19}, is based on the established fact \cite{r20} that RT
ferromagnetism can only be achieved when $\Delta$E is larger than
about 30 meV. Note again that for the ZnO:Gd system, the \emph{s-f}
coupling is much larger than the \emph{f-f} and \emph{f-p}
couplings, which makes it be the main factor for the
electron-mediated ferromagnetism in ZnO:Gd system.

For ZnO:Nd system, the holes can also mediate the ferromagnetism
according to our calculated results listed in table I. In the
neutral case, the calculated energy difference $\Delta$E for the
ZnO:Nd system is 94 meV. In the hole-doped cases, the energy
differences $\Delta$E are 128 and 134 meV for 0.25 and 0.50 hole per
Nd, respectively. Thus the FM coupling with low or intermediate hole
concentration is more stable than that of the neutral system.
However, when the concentration of holes is as high as 0.75 (1.0)
hole per Nd, the energy difference $\Delta$E is 69 (20) meV, which
indicates that overdoping of holes to the ZnO:Nd system will lower
$\Delta$E and the subsequent stability of the FM phase. For electron
doping, according to our calculated $\Delta$E in Table I, the
stability of FM ordering of ZnO:Nd system is weakened. In general,
grown ZnO sample is always \emph{n}-type conducting. Based on our
calculated results, if the electron concentration is relatively low
(0.25 electron/Nd for example shown in Table I), the RT
ferromagnetism can also be achieved for ZnO:Nd system. Ungureanu
\emph{et al.} \cite{r19} also reported that the ferromagnetism is
present in ZnO:Nd system similar to that of ZnO:Gd system at the
same doping level. For vacancies, both V$_{\rm{Zn}}$ and
V$_{\rm{O}}$ in ZnO:Nd system result in the FM phase with
$\Delta$E=52 and 67 meV, respectively.

Due to the strong on-site Coulomb repulsion among the localized
4\emph{f} electrons, the traditional DFT with GGA (PBE) or LDA
scheme can not accurately described the strong correlation.
Therefore, several methods are adopted to overcome the drawback
mentioned above, such as LDA+\emph{U}\cite{r21}, and
Heyd-Scuseria-Ernzerhof (HSE) hybrid-functional\cite{r22,r23}. For
GaN:Gd system, Dalpian and Wei \cite{r6} also performed LDA+\emph{U}
calculation to test the adequacy of the LDA and their results show
that the qualitative picture is unchanged besides some quantitative
changes. Therefore, we may expect that the main trends of carries
mediated ferromagnetism discussed in our paper will not change if
the Hubbard \emph{U} correlation is taken into account.

\section{\label{sec:level1}SUMMARY}
In summary, we have systematically investigated the magnetic
properties of ZnO:RE (RE=Gd, Nd) systems in different charged
states. Because of the \emph{s-f} coupling between the Gd ions
\emph{f} and the host \emph{s} states, the electrons can mediate the
ferromagnetism in the ZnO:Gd system. We present that the RT
ferromagnetism can be achieved in ZnO:Gd system if electrons of
appropriate concentration are doped in the sample. However, for
ZnO:Nd system, the holes can enhance the stability of the FM
ordering, and the RT ferromagnetism can also be achieved in the
\emph{n}-type ZnO:Nd. Our calculated results agree well with the
recent experimental observation.

\section{\label{sec:level1}ACKNOWLEDGMENTS}

This work was supported by the National Basic Research Program of
China (973 Program) grant No. G2009CB929300 and the National
Natural Science Foundation of China under Grant Nos. 60521001 and
60776061.

\end{document}